\documentclass[fleqn,10pt]{wlscirep}
\usepackage[utf8]{inputenc}
\usepackage[T1]{fontenc}

\usepackage{graphicx}
\usepackage{amsmath}
\usepackage{amsfonts}
\usepackage{amssymb}
\usepackage{hyperref}
\usepackage{braket}
\usepackage{xcolor}
\usepackage{multirow}
\usepackage{caption}
\usepackage{subcaption}

\newcommand{\Ab}{A_{\text{big}}}
\newcommand{\As}{A_{\text{small}}}

\newcommand{\cross}[1][0.5pt]{{\ooalign{\rule[1ex]{1ex}{#1}\cr\hss\rule{#1}{0.7em}\hss\cr}}}

\title{Quantum Computing and Preconditioners for Hydrological Linear Systems}

\author[1,*]{John Golden}
\author[2]{Daniel O'Malley}
\author[2]{Hari Viswanathan}
\affil[1]{CCS-3 (Information Sciences), Los Alamos National Laboratory, Los Alamos, 87545, USA}
\affil[2]{EES-16 (Computational Earth Sciences), Los Alamos National Laboratory, Los Alamos, 87545, USA}

\affil[*]{golden@lanl.govk}

\begin{abstract}
Modeling hydrological fracture networks is a hallmark challenge in computational earth sciences. Accurately predicting critical features of fracture systems, e.g. percolation, can require solving large linear systems far beyond current or future high performance capabilities. Quantum computers can theoretically bypass the memory and speed constraints faced by classical approaches, however several technical issues must first be addressed. Chief amongst these difficulties is that such systems are often ill-conditioned, i.e. small changes in the system can produce large changes in the solution, which can slow down the performance of linear solving algorithms. We test several existing quantum techniques to improve the condition number, but find they are insufficient. We then introduce the inverse Laplacian preconditioner, which improves the scaling of the condition number of the system from $O(N)$ to $O(\sqrt{N})$ and admits a quantum implementation. These results are a critical first step in developing a quantum solver for fracture systems, both advancing the state of hydrological modeling and providing a novel real-world application for quantum linear systems algorithms. 
\end{abstract}
\begin{document}

\flushbottom
\maketitle
\thispagestyle{empty}

\section{Introduction}

Quantum algorithms for solving linear systems offer a potential for a practical quantum advantage over classical algorithms~\cite{harrow2009quantum}.
However, several technical and conceptual challenges remain before a meaningful, real-world example of quantum advantage can be exhibited for linear systems.
On the technical side, many of these quantum linear systems (QLS) algorithms can only be implemented on future error-corrected hardware~\cite{harrow2009quantum}, while others are designed for today's noisy intermediate-sized quantum devices but feature a less pronounced speed-up~\cite{bravo2019variational}.
On the conceptual side, existing algorithms feature numerous caveats that must be satisfied in order for good performance~\cite{aaronson2015}.
The ideal system $Ax=b$ for exhibiting quantum speed-up satisfies:
	1) $A$ is well-conditioned,
	2) the quantum operator $e^{-iAt}$ can be efficiently implemented,
	3) a quantum state proportional to $b$ can be efficiently prepared,
	4) only specific values or statistics of the solution $x$ are of interest,
	5) there does not exist a classical algorithm which exploits these or other properties of the system to solve it equally as fast.
The final point is particularly notable, as large systems of equations can often be replaced by small systems of equations which retain sufficient accuracy (and can be solved classically with modest cost)~\cite{montanaro2016quantum}.
Existing applications for QLS algorithms have therefore been largely synthetic or specific examples, carefully chosen to avoid these constraints but also lacking in real-world application~\cite{huang2019near,dukalski2021}.

In this work we initiate the study of simulating fluid flow through fracture systems with QLS algorithms.
Hydrological flow is a very challenging problem in geophysics, because the contrast between the scales on which simulations are often done (kilometers or larger) and the scale of heterogeneities (centimeters or smaller) requires discretizing the problem over very large meshes.
Here we address the common problem of determining the pressure of a subsurface liquid (e.g. water or oil), either at a specific location, i.e. at a well, or averaged across a wide region, i.e. effective permeability.
Extracting the pressure at an individual location or studying averaged properties (rather than needing exact pressure at every point in the system) addresses point 4 in our list of QLS requirements.
As we argue in Sec.~\ref{sec:coarse-grain}, this type of problem cannot be reduced to a smaller system of equations without losing critical information, and so current classical techniques cannot capture the full scope of the problem at real-world scale (point 5).
Since the matrices $A$ appearing in subsurface flow applications are generally sparse, and the vectors $b$ are relatively uniform, points 2 and 3 are plausibly addressed via quantum RAM~\cite{aaronson2015}.
However, these points are quite nuanced and will be studied in detail in a future work. 

Our focus for this work is point 1, improving the condition number of the linear system.
The condition number of a matrix describes how sensitive the system is to small perturbations, or more specifically, how much an error in the right hand side, $b$, propagates to an error in the solution, $x$, in the worst case.
Easy matrices, such as unitary matrices, have condition number 1 and increasing the condition number makes the linear system harder to solve.
Both quantum and classical linear systems algorithms generally perform worse on systems with large condition numbers.
Classical methods to reduce the condition number, known as preconditioners, have been successfully developed and tailored specifically for fracture systems, the most prominent example being multigrid preconditioners~\cite{greer2022comparison}.
However, they cannot simply be ported directly onto quantum computers due to the drastically different underlying technologies and algorithmic design constraints~\cite{asiani2021}.
Therefore, in order to use quantum computers to solve large linear systems for fracture networks of relevance and outpace classical techniques, an effective quantum preconditioning algorithm for fracture networks must be developed.
A good preconditioner is generally tied to the type of linear system being solved and exploits some underlying mathematical or physical structure in the problem.
Meanwhile, one must also find an efficient way of calculating the preconditioned form of the system, which will be necessary to do on the quantum computer itself for very large systems due to memory constraints.
These two constraints -- tailored to a specific application while also having an efficient quantum implementation -- are likely to play a prominent role in future attempts to solve large, interesting linear systems on quantum computers.

\begin{figure}[t!]
	\centering
	\begin{subfigure}[b]{0.8\textwidth}
		\begin{tabular}{lcl} 	
			\textbf{Common QLS Caveat for $Ax=b$} && \textbf{Resolution for Subsurface Flow} \\
			1. $A$ is well-conditioned &$\rightarrow$& inverse Laplacian preconditioner (see Fig.~\ref{fig:results}b and Sec.~\ref{sec:test_precon}) \\
			2. efficiently implement $e^{-iAt}$ &$\rightarrow$& future work \\
			3. efficiently prepare $\ket{b}$ &$\rightarrow$& future work \\
			4. don't need full $x$ &$\rightarrow$& interested in average pressure, or at specific locations (e.g. wells)\\
			5. no classical counterpart &$\rightarrow$& memory needs, can't coarse grain (see Sec.~\ref{sec:coarse-grain})
		\end{tabular}
		\caption{}
	\end{subfigure}
	\hfill
	\begin{subfigure}[b]{0.7\textwidth}
		\includegraphics[width=\textwidth]{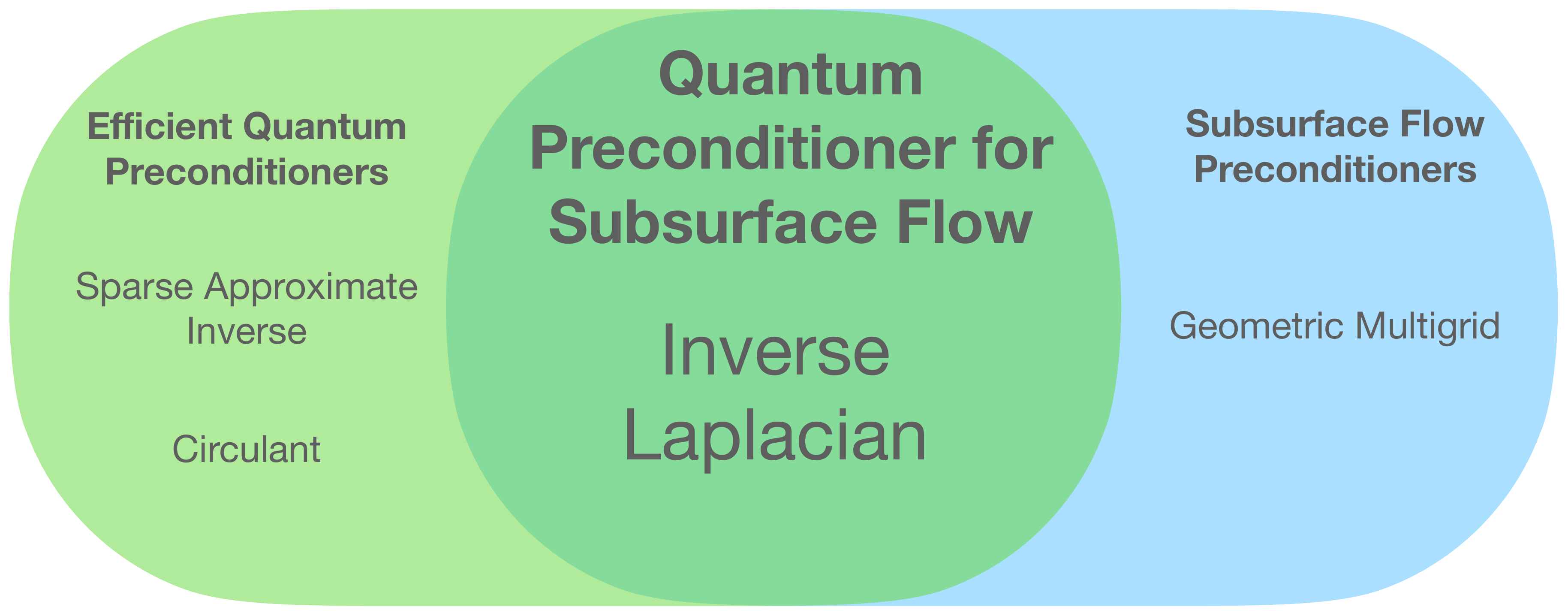}
		\caption{}
	\end{subfigure}
	\caption{Summary of results. \textbf{(a)}: Common requirements for good performance from QLS algorithms, and the ways in which subsurface flow problems can be made to meet them. Our primary contributions in this work are points 1 and 5, namely, introducing the inverse Laplacian preconditioner and providing arguments against the scalabality of classical methods.\\ \textbf{(b)}: Existing quantum preconditioners are not effective on subsurface flow through fracture systems, while classical fracture preconditioning techniques cannot reasonably be enacted via quantum algorithms. The inverse Laplacian is well-suited to fracture systems while also being efficiently realized on a quantum computer.}
	\label{fig:results}
\end{figure}

We show that existing techniques to reduce the condition number either do not have quantum implementations or are not well-suited for fracture systems.
However, we find that the inverse Laplacian is both effective on fracture systems and can be efficiently implemented on a quantum computer. 
See Fig.~\ref{fig:results} for a summary of our results.
Furthermore, we show that realistic fracture systems can be solved by a quantum computer with a polynomial advantage over classical approaches.
Finally, we discuss the remaining algorithmic bottlenecks which need to be resolved to unlock the full potential of QLS for fracture systems.

\section{Results}\label{sec:results}
In this section, we describe the geophysical problem in further detail and argue why it is not fully tractable for classical computers (Sec.~\ref{sec:coarse-grain}).
We then review the three existing quantum preconditioner algorithms and study how well they improve the condition number of several synthetic fracture network examples.
The condition number (without preconditioning) of these examples all scale as $O(N)$, and the best of the preconditioners reduces this scaling to roughly $O(\sqrt{N})$ (Sec.~\ref{sec:test_precon}), which is potentially sufficient to produce a polynomial advantage over classical techniques.
Finally, we analyze how incorporating the preconditioner into a full QLS algorithm might affect overall runtime (Sec.~\ref{sec:scaling}). 

\subsection{Fracture Networks and Coarse-Graining}\label{sec:coarse-grain}
\begin{figure}[t!]
	\begin{centering}
	\includegraphics[width=\textwidth]{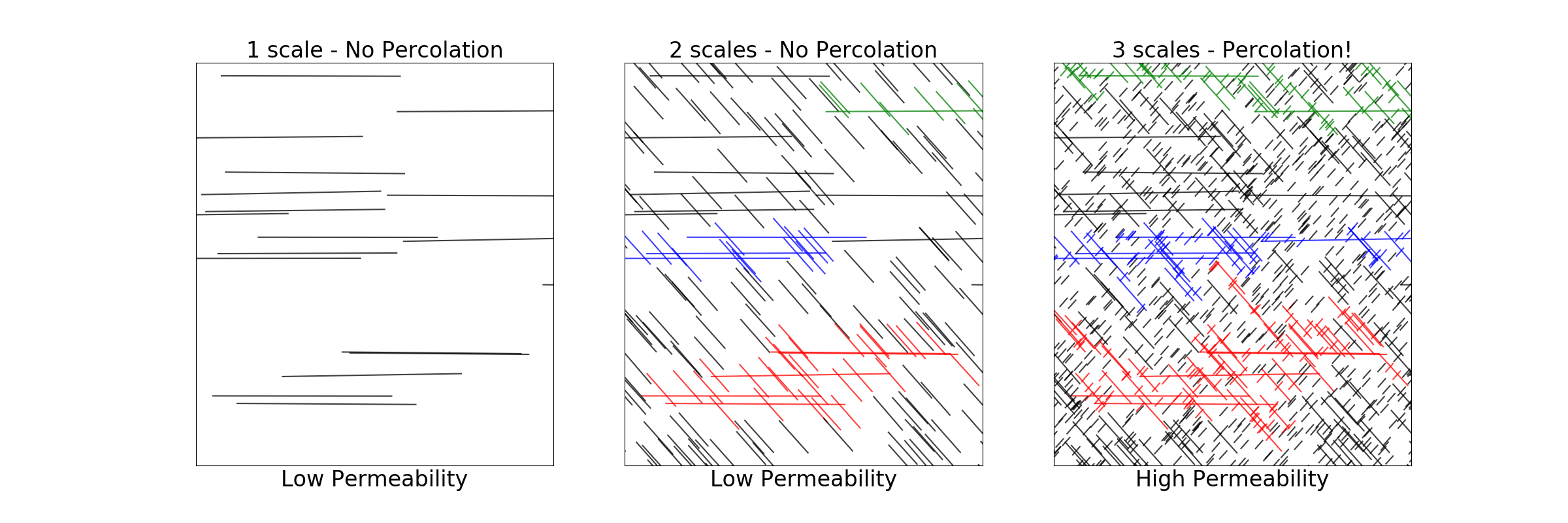}
	\caption{Fracture systems feature critical behavior, such as percolation, which is only apparent when taking many length scales in to account. Here we give a simplified conceptual diagram of the heirarchical structure of fracture networks. This model highlights how percolation, which has a large impact on effective permeability, is the result of many fractures of different lengths intersecting (here all fractures are represented as lines in two dimensions and we do not explicitly model the width or aperture). The red/blue/green colors indicate connected components in the fracture network -- the blue connected component in the right image results in percolation.}
	\label{fig:percolation}
	\end{centering}
\end{figure}

In a complex fracture network, fractures of many scales -- from kilometers to centimeters -- intersect. 
Critically, small fractures cannot generally be neglected because these can transform the network topology radically, e.g. pushing a system over a percolation threshold, see Fig.~\ref{fig:percolation}.
Small fractures may also collectively contribute a large surface area to the network providing a critical connection between fractures and the underlying rock.
When modeling flow in these networks, it is therefore critical to include the full range of fracture scales, which has led to the development of advanced meshing techniques~\cite{hyman2015dfnWorks} and high-performance simulators~\cite{mills2007simulating}.
However, these approaches do not provide a viable path to modeling the full range of scales.
Even state-of-the-art high-performance computers and cutting-edge methods can only model large fracture networks with fracture lengths varying over three orders of magnitude~\cite{hyman2016fracture}.
Modeling real-world fracture networks to a high degree of accuracy requires meshes far beyond current or future classical capabilities—e.g., a 1km domain with a 1cm resolution would require $10^{15}$ degrees of freedom.
Larger systems of equations would be needed for larger domains or more finely resolved meshes.

Numerical models of subsurface flow in a fracture network are based on a discretized version of \cite{greer2022comparison}
\begin{equation}
	\nabla \cdot(k\nabla h)=f
	\label{eq:groundwaterflow}
\end{equation}
where $k$ is the permeability, $f$ is the fluid flux, and $h$ is the pressure.
Each of $k$, $f$, and $h$ is a funciton that varies spatially throughout the domain.
The pressure is key for applications of subsurface flow such as waste fluid disposal and hydraulic fracturing \cite{pachalieva2022physics}.
It is also critical for transport applications, since pressure gradients drive the transport.
In fractured systems, $k$ is highly heterogeneous with a sharp increase when moving from the rock matrix (where $k$ is small) to a fracture (where $k$ is large).

We discretize Eq.~\ref{eq:groundwaterflow} using a two-point flux finite volume method, which is one of the standard numerical schemes used in subsurface flow solver code bases including: {\sc fehm}~\cite{zyvoloski2007fehm}, {\sc tough2}~\cite{pruess1999tough2}, and {\sc pflotran}~\cite{lichtner2015pflotran}. 
The two-point flux finite volume method ensures mass conservation, which is a highly desireable property for these numerical solvers.
This results in a coefficient matrix which is sparse, symmetric, and positive definite.
We treat the permeability of the rock matrix as being constant.
While the approximation is imperfect, it is a major step up from discrete fracture networking approaches which effectively treat the rock matrix as having zero permeability \cite{greer2022comparison}.
This approach does, however, capture the basic physics of fracture flow -- most but not all of the flow occurs in the high-permeability fractures.

\begin{figure}[t!]
	\begin{centering}
	\includegraphics[width=0.9\textwidth]{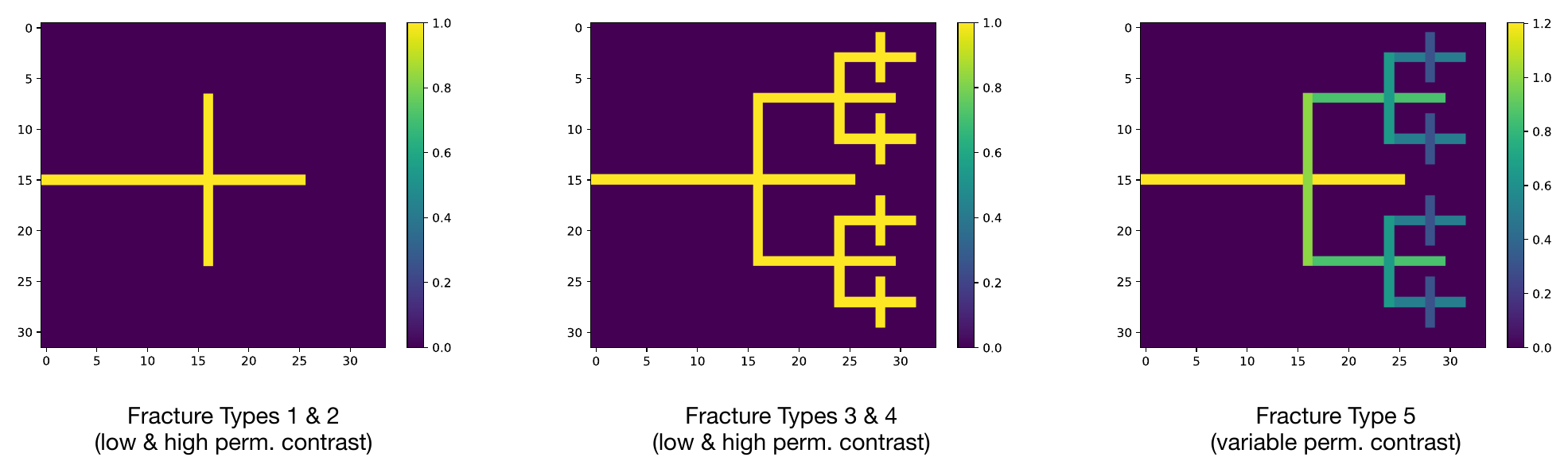}
	\caption{2D fracture networks under consideration. On the left, Fracture Types 1 \& 2 feature a simple \cross-style fracture network (with low \& high permeability contrast, respectively). In the middle, Fracture Types 3 \& 4 add more fractures via fractal recursion, keeping the permeability constant across fractures (again with low \& high permeability contrast, respectively). On the right, Fracture Type 5 gives less permeability to shorter fractures.}
	\label{fig:fracture_imgs}
	\end{centering}
\end{figure}

In this work, we study a variety of 2D fracture network models.
The simplest system we studied involved two fractures intersecting in a \cross-configuration, and we then studied fractal-style recursion of the \cross-system to generate more complicated fracture networks, see Fig.~\ref{fig:fracture_imgs}.
The relative permeability of the fractures as compared to the underlying rock is a critical parameter in the analysis of fracture systems, and we studied five different types:
\begin{enumerate}[leftmargin=3cm]
	\item[Fracture Type 1:] ``Simple, Low'' is the \cross-system with fracture 10\% more permeable as the underlying rock.
	\item[Fracture Type 2:] ``Simple, High'' is the \cross-system with fractures $10^4$ times more permeable than the underlying rock.
	\item[Fracture Types 3 \& 4:] ``Fractal, Low/High'' are the same as above, but with the fractal system.
	\item[Fracture Type 5:] ``Fractal, Var.'' is the fractal system where the fractures have permeability contrast that scales down as the fractures get smaller, i.e. largest fractures have contrast $10^4$ and the smallest fractures have contrast $1.1$, with the contrast scaling down as the $(\text{fracture length})^{1/2}$, commonly used in pracice~\cite{hyman2016fracture}.
\end{enumerate}

\subsection{Test of Existing Quantum Preconditioners}\label{sec:test_precon}

Recently developed QLS algorithms provide a novel path to modeling the full complexity of fracture networks.
Since a quantum computer with $n$ qubits can represent a $2^n$ dimensional vector, vast systems of equations can be solved with a small number of qubits.
That is, for the 1km domain with a 1cm resolution problem described above, a quantum computer could require as few as $O(\log(10^{15})\approx50)$ qubits, whereas a classical computer would require $O(10^{15})$ classical bits.
While additional quantum resources will likely be necessary, either in the form of ancilla qubits or quantum RAM, this highlights the raw potential of the quantum approach for the huge systems of equations necessary to accurately model realistic subsurface flow.
Furthermore, the computational complexity of quantum linear systems algorithms can in some cases be exponentially better than the best classical counterparts~\cite{childs2017quantum}.

The original QLS algorithm introduced by Harrow, Hassidim, and Lloyd (HHL)~\cite{harrow2009quantum} solves a sparse $N$-variable system of equations $Ax = b$ with a runtime of $O(\log(N) \kappa(A)^2)$, where $\kappa(A) := \|A\|\|A^{-1}\|$ is the condition number of $A$ (in this work we use $\|A\|$ with no subscripts to refer to the 2-norm (or operator norm), i.e. $\|A\|\equiv \|A\|_2$, and explicitly use $\|A\|_F$ to refer to the Frobenius norm). 
The best classical algorithm, the Conjugate Gradient method, runs in $O(N\sqrt{\kappa(A)})$ on sparse matrices, so the quantum algorithm provides an exponential speed-up when $\kappa(A)$ is small (for notational clarity, we often use `$\kappa$' alone to signify $\kappa(A)$).

Since the introduction of the HHL algorithm, many more QLS algorithms have been introduced, including improvements to HHL~\cite{childs2017quantum,ambainis2010variable}, adiabatic approaches~\cite{subasi2019quantum,costa2021optimal}, and variational algorithms implementable on near-term quantum computers~\cite{huang2019near,bravo2019variational}.
A feature common to all of these quantum approaches is a scaling with $\kappa$ that is linear at best, as compared to the $O(\sqrt{\kappa})$ scaling of the best classical approaches.
A common technique in classical analysis is to further reduce $\kappa$ by using a preconditioner.
This technique relies on finding a matrix $M$ such that $\kappa(MA) \ll \kappa(A)$.
One then finds the $x$ satisfying $MAx = Mb$.
The matrix $M$ is generally dependent on the specific matrix $A$, and different preconditioning approaches have been developed for different contexts.

Despite the significant interest and activity in QLS algorithms, relatively little work has been done to develop application-specific preconditioning algorithms.
It is important to emphasize that in this work we only consider preconditioning algorithms which are implementable on a quantum computer, as opposed to any sort of hybrid classical-quantum preconditioning method.
While we do not rule out the possibility of such an approach, the extreme memory requirements of full-scale subsurface flow problems (as described above) suggests that calculating the preconditioned matrix classically would be intractable.
There are currently three general purpose quantum preconditioning algorithms in the literature: the circulant method~\cite{shao2021quantum}, the sparse approximate inverse method~\cite{clader2013preconditioned}, and the fast-inverse method~\cite{tong2021fast}.
These algorithms are described in detail in Sec.~\ref{sec:methods}, here we give only the salient points.

The circulant method is a one-size-fits-all approach, that is, the only input is a matrix $A$ and the output is a preconditioner $M$.
With SPAI one gives a sparsity pattern for the preconditioner $M$, and several techniques have been developed for determining good sparsity patterns for fracture systems~\cite{labutin2013algorithm}.
The fast-inverse method is designed for systems of the form $A+B$, where $A^{-1}$ can be easily calculated.
One then uses $A^{-1}$ as the preconditioner.
As discussed in Sec.~\ref{sec:methods}, fracture systems can be decomposed into $\Delta + A_F$, where the Laplacian $\Delta$ describes the system in the absence of fractures, and $A_F$ is the contribution of the fractures.
Because the singular value decomposition of the Laplacian is known~\cite{demmel1997applied}, $\Delta^{-1}$ can be efficiently calculated and used as a preconditioner.

\begin{figure}[t!]
	\begin{centering}{}
	\includegraphics[width=\textwidth]{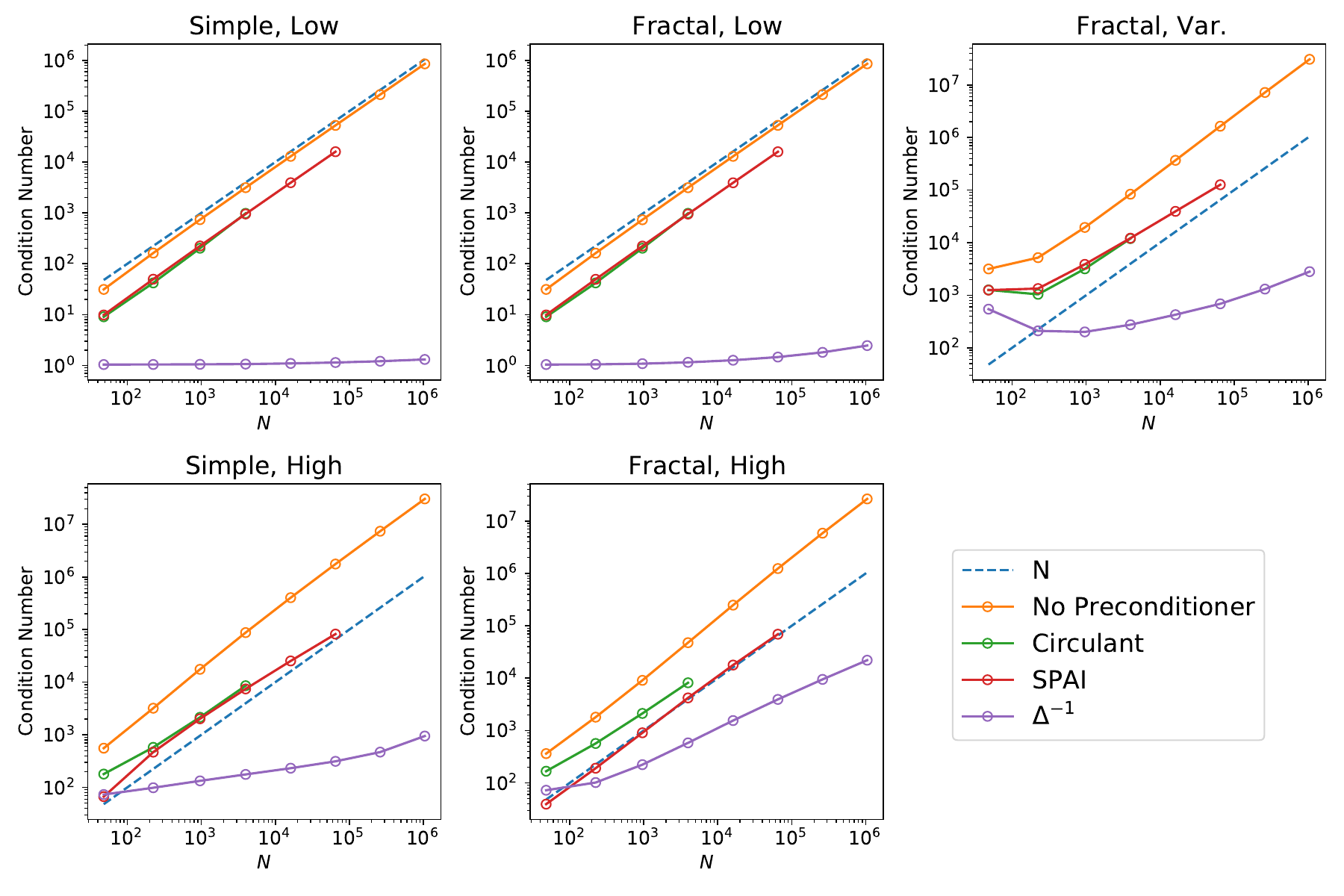}
	\caption{Efficacy of the preconditioners under analysis. The inverse Laplacian $\Delta^{-1}$ gives the best scaling in all cases, while the circulant and SPAI preconditioners reduce $\kappa(A)$ but do not significantly improve the scaling in $N$. There are fewer data points for the circulant and SPAI preconditioners due to computational constraints.}
	\label{fig:precon_scaling}
	\end{centering}
\end{figure}

In Fig.~\ref{fig:precon_scaling} we numerically evaluate the effect of the circulant, SPAI, and inverse Laplacian preconditioners on all of the fracture types described in Sec.~\ref{sec:coarse-grain} up to $O(10^6)$ variables.
To estimate how the preconditioner performance scales with $N$, we perform a linear regression on the logarithm of the final four data points for each preconditioner applied to each fracture type.
We do not include the first points as they sometimes exhibited variance due to small matrix sizes, however for $N>10^3$, all of the results show clear exponential scaling in the number of qubits and polynomial scaling in the number of equations.
We find that the circulant and SPAI methods are poor choices for the fracture systems under consideration.
While these two preconditioners consistently reduce the condition number of the system, they do not improve how the condition number scales in $N$, which is necessary to unlock the full potential of QLS algorithms for large problems.
The inverse Laplacian preconditioner, however, does meaningfully improve the scaling of the condition number.
In the cases with low permeability contrast, the condition number of the system $\Delta^{-1}A$ is very low, scaling as $\le O(N^{0.05})$.
The high permeability contrast systems do not fair as well, with the $\kappa$ of the preconditioned fractal system scaling as $O(N^{0.6})$.
The fractal system with variable permeability, which is the most realistic of the systems under consideration, has a preconditioned $\kappa$ which asymptotically scales as approximately $O(N^{0.55})$.

\subsection{Algorithmic Scaling with Inverse Laplacian Preconditioner}\label{sec:scaling}

Identifying a preconditioner $M$ which reduces the condition number of a system $A$ is generally not sufficient to guarantee good performance of a QLS algorithm to solve the preconditioned system $MAx = Mb$. 
This is because one must find an efficient way to calculate the product $MA$ in such a way as to make it readily accessible on a quantum computer.
This can be accomplished either through a classical oracle which efficiently calculates $MA$, or through a quantum algorithm which uses oracles for $A$ and $M$ and then calculates $MA$.
As described previously, in this work we only study cases where the preconditioner can be applied on the quantum computer itself, due to the considerable memory demands of full-scale hydrological simulations.
Unfortunately, simply calculating $MA$ in the general case on a quantum computer with generic matrix multiplication adds an $O(N^2)$ overhead~\cite{shao2018quantum} and would remove any benefit from the reduced condition number.
Each of the three methods previously discussed get around this limitation through clever techniques: the circulant method calculates the product with the quantum Fourier Transform algorithm, the SPAI method exploits sparseness of $M$ and $A$, and the fast-inverse method assumes efficient block-encodings of $M$ and $A$. 

As described in Sec.~\ref{sec:test_precon}, the data presented in Fig.~\ref{fig:precon_scaling} shows that the circulant and SPAI preconditioners do not meaningfully improve the scaling in $N$ of the condition number for the fracture examples considered in this paper.
Therefore, it is not worth determining the full algorithmic runtime for implementing either of these preconditioners in a full QLS algorithm.
However, the inverse Laplacian reduces the condition number to scale between $O(N^{0.05})$ to $O(N^{0.6})$, which is potentially sufficient for an advantage over classical algorithms. 
It is thus instructive to estimate the impact the inverse Laplacian preconditioner has on the overall runtime for solving subsurface flow systems.

While implementing the inverse Laplacian preconditioner into the original HHL algorithm is potentially possible, the QLS algorithm of Tong \emph{et al.}~\cite{tong2021fast} already gives a direct method of applying the preconditioner and solving the resulting system.
We can therefore use the scaling of their algorithm as a proof-of-concept to gain an understanding of whether the inverse Laplacian improves the condition number sufficiently to recover some quantum speed-up.
This analysis purposefully ignores intricacies resulting from points 2 and 3 in our list of QLS caveats, i.e. efficiently turning the classical data $A$ and $b$ into appropriate quantum states and operators.
We emphasize that this analysis is intended as a conservative estimate of how a preconditioned QLS algorithm might scale when solving fracture systems.
There are hopefully more efficient implementations, which we discuss more in Sec.~\ref{sec:discussion} and will explore further in future work.

\begin{figure}[t!]
	\begin{centering}
	\includegraphics[width=0.75\textwidth]{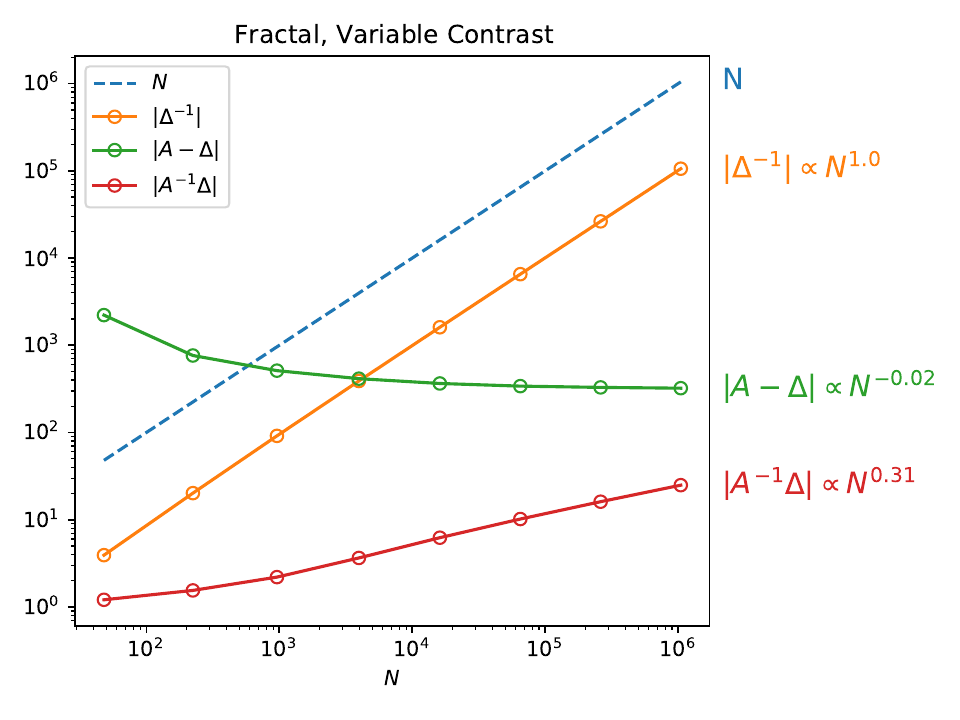}
	\caption{Scaling of the components of the fast-inverse QLS algorithm for problems of increasing size for a fractal fracture network with high permeability contrast.}
	\label{fig:fastinv_scaling}
	\end{centering}
\end{figure}

As we show in Sec.~\ref{sec:methods}, the fast-inverse QLS algorithm with $\Delta^{-1}$ as the preconditioner gives a runtime bounded below by
\begin{equation}
	O(\|\Delta^{-1}\|\|A-\Delta\|\|A^{-1}\Delta\|\log(\|A^{-1}\Delta\|)/\epsilon).
\end{equation}
The scaling in $N$ of $\|A-\Delta\|$ and $\|A^{-1}\Delta\|$ are dependent on the exact fracture systems being studied and must be determined experimentally.
In Fig.~\ref{fig:fastinv_scaling} we show the scaling of these components for the fractal system with variable contrast.
We focus on this particular example since it is the most realistic of the different fracture types.
As was the case in Sec.~\ref{sec:test_precon}, we estimate the large-$N$ scaling of the different components by linear regression (on the log-log plot) of the data points for $N>10^3$.
In Table~\ref{tab:fastinv} we summarize the scaling of each component as well as the overall scaling (modulo $\log(N)$) compared with the scaling for Conjugate Gradient on the same systems.

\begin{table}
\centering
\begin{tabular}{c|c|c|c|c||c||c|}
\multicolumn{1}{r}{} & \multicolumn{6}{c}{Scaling in $N$ of:} \\
  & \multicolumn{1}{c}{$\kappa(A)$} & \multicolumn{1}{c}{$\|\Delta^{-1}\|$} & \multicolumn{1}{c}{$\|A-\Delta\|$} & \multicolumn{1}{c}{$\|A^{-1}\Delta\|$} & \multicolumn{1}{c}{$\widetilde{\text{Overall}}$} & \multicolumn{1}{c}{Classical} \\ \hline
Simple, Low Contrast & 1 & 1.01 & 0 & 0.03 & 1.03 & 1.5\\ \hline
Simple, High Contrast & 1.02 & 1.01 & 0 & 0.26 & 1.26 & 1.51\\ \hline
Fractal, Low Contrast & 1 & 1.01 & 0 & 0.11 & 1.11 & 1.5\\ \hline
Fractal, High Contrast & 1.09 & 1.01 & 0 & 0.34 & 1.34 & 1.54\\ \hline
Fractal, Var. Contrast & 1.05 & 1.01 & -0.02 & 0.31 & 1.3 & 1.52\\ \hline
\end{tabular}
\caption{Scaling of the various components entering the overall scaling of the fast-inverse QLS using $\Delta^{-1}$ as the preconditioner. The $\widetilde{\text{Overall}}$ scaling reported is modulo $\log(N)$.}
\label{tab:fastinv}
\end{table}

Using the fast inverse QLS algorithm with the inverse Laplacian preconditioner, we can potentially achieve a polynomial improvement over the best generic classical scaling for all fracture systems considered here.
This approach utilizes block encodings of $\Delta^{-1}$ and $A-\Delta$ to calculate the product $\Delta^{-1}A$.
However, since $\|\Delta^{-1}\|$ scales linearly in $N$, the block encoding takes at least this long.
Future algorithms for QLS, specifically tailored to fracture systems, could be developed to calculate $\Delta^{-1}A$ even more efficiently by exploiting the sparseness of $A$.

\section{Discussion}\label{sec:discussion}

In this work, we have initiated the study of using QLS algorithms to solve systems of equations describing subsurface flow.
As we argue in Sec.~\ref{sec:coarse-grain}, capturing the full behavior of these systems at real-world scale is prohibitively memory-intensive for classical computers.
Quantum computing provides an alternative path, potentially using significantly less resources and offering improved scaling in problem size.
However, several conceptual problems must be addressed (in addition to the need for improved quantum computing hardware).
The most prominent of these issues are the poor condition number of these systems and the means of loading information onto the quantum computer.
Here we have studied the first problem through the use of quantum preconditioners, and we leave the latter problem to future work. 

We have shown that two previously introduced quantum preconditioners, the circulant and SPAI methods, do not improve the scaling in $N$ of $\kappa(A)$ and therefore will not help gain a quantum advantage for these fracture systems.
However, the inverse Laplacian is an effective preconditioner for fracture systems and readily admits a quantum implementation.
In particular, it can be implemented via the fast-inverse QLS algorithm, and the overall scaling of this solver scales better than the best generic classical algorithm.

In comparing against classical techniques, we have so far not addressed the fact that for PDE-based systems on a uniform mesh, such as those considered here, more specialized methods can be used.
Geometric multigrid methods, which exploit the structured mesh, can solve systems of equations in $O(N)$ or $O(N\log N)$~\cite{strang2007computational}.
Due to the extensive caveats and stringent requirements attached to QLS algorithms, it is noteworthy (though by no means sufficient) that our preliminary results suggest that quantum performance will be at least comparable to state-of-the-art, highly tuned classical techniques.
This is in addition to the memory requirements which classical techniques inevitably hit.
Still, it is clearly necessary to further refine QLS algorithms with an even greater eye towards the specific physics and mathematical structures of the fracture systems at hand.

An obvious area of improvement is a more efficient quantum means of implementing the preconditioned system $\Delta^{-1}A$.
In the most realistic case we studied, the fractal system with variable contrast, $\kappa(\Delta^{-1}A)$ asymptotes to roughly $O(N^{0.55})$. 
Therefore in principle the scaling of solving just the preconditioned system, e.g. with recent adiabatic QLS algorithms~\cite{subasi2019quantum}, would be $O(N^{0.55}\log{N})$, a significant improvement over the geometric multigrid methods.

Alternatively, while the inverse Laplacian opens the door for polynomial speed-up over classical, an even better preconditioner is needed for exponential speed-up.
For example, while a direct quantum port of the classical multigrid methods is not plausible, some of the ideas may be used to construct an analagous approach that can be implemented on a quantum computer.
Reducing the condition number scaling to $O(\log{N})$ in the quantum context would then be possible.

This study shows that fracture networks are a challenging real-world problem with a potential for serious advancements from quantum computation.
The large linear systems necessary to accurately model flow behavior are sparse yet cannot be coarse-grained.
The condition number $\kappa$ of the systems tends to scale linearly with $N$, however the inverse Laplacian preconditioner, which readily admits a quantum implementation, can improve this scaling considerably, and it is likely evern further advancements can be made in $\kappa$.
Future work will be devoted to incorporating our application-specific preconditioning techniques into a full quantum linear solver, ideally targeting an implementation on NISQ devices.

\section{Methods}\label{sec:methods}

In this section we provide a more detailed review of the preconditioner methods evaluated in Sec.~\ref{sec:test_precon}, as well as a derivation of the scaling bound for the fast-inverse algorithm found in Sec.~\ref{sec:scaling}.

\subsection{Circulant Preconditioner}
The circulant preconditioner method of Shao \emph{et al.}~\cite{shao2021quantum} gives an efficient quantum implementation of a circulant preconditioner $C$ based on the quantum Fourier transform $F$.
An $n\times n$ matrix $C$ is circulant if $C_{ij} = C_{(i-j)\mod n}$.
The use of circulant preconditioners in classical applications is motivated by the fact that, for a given circulant matrix $C$ and an arbitrary matrix $A$, $CA$ and $C^{-1}A$ can be computed in $O(n \log n)$ steps using the fast Fourier transform.
Circulant preconditioners are particularly useful in solving Toeplitz systems~\cite{gray2006toeplitz}. 

For an arbitrary matrix $A$, one can construct the circulant preconditioner via
\begin{equation}
	C(A) = F^{\dagger}\text{diag}(FAF^{\dagger})F,
\end{equation}
where $F_{jk} = \frac{1}{\sqrt{n}}\omega^{jk}$ with $\omega = e^{-2\pi i/n}$.
$C^{-1}(A)$ is then used the preconditioner.
$F$ can be efficiently implemented via the quantum Fourier transform, and the middle term simplifies to 
\begin{equation}\label{eq:circulant}
	\text{diag}(FAF^{\dagger})_k = \frac{1}{n}\sum_{p,q}\omega^{(p-q)k}A_{p,q}.
\end{equation}
An algorithm for efficiently preparing the state in Eq.~\ref{eq:circulant} is given in~\cite{shao2021quantum}.
This approach works for arbitrary dense non-Hermitian matrices, however there is no upper bound on $\kappa(CA)$, and in practice for random dense matrices $\kappa(CA) = O(\kappa(A))$.

\subsection{Sparse Approximate Inverse}
The Sparse Approximate Inverse (SPAI) approach for solving a system $Ax=b$ attempts to find a matrix $M$ such that $MA \approx I$, where $M$ has a (user-defined) sparsity pattern.
For example, if one gives a sparsity pattern involving $n$ non-zero rows and $d$ non-zero elements per row, then $M$ is given by solving $n \times d$ independent least squares problems.
The trick with this approach is determining which sparsity pattern to choose for $M$.

Clader \emph{et al.}~\cite{clader2013preconditioned} showed that the preconditioned system
\begin{equation}\label{eq:spai_eq}
	MAx = Mb
\end{equation}
can be solved via a slightly modified version of the HHL algorithm.
The overall scaling for actually solving Eq.~\ref{eq:spai_eq} with error $\epsilon$ is 
\begin{equation}
	\tilde{O}(d^7 \kappa(MA)\log(N)/\epsilon^2).
\end{equation}

In Sec.~\ref{sec:test_precon} we adopt the relatively standard approach of using the sparsity pattern of $A$ for $M$.
One can also try other methods~\cite{labutin2013algorithm}, which can significantly reduce the condition number, but again do not improve the scaling in $N$ for the fracture systems studied here.
Finally, $d$ does not need to be a constant.
As long as $\kappa(MA) = O(1)$, the sparsity pattern can scale with $d \propto N^{\le1/7}$ in order to at least recover some quantum advantage.
However, for the systems and sparsity patterns considered here, a small increase in $d$ has a corresponding small decrease in $\kappa(MA)$.
For example, when applying the technique described in~\cite{labutin2013algorithm} to the ``simple'' systems, i.e. fractal depth 1, increasing the density of $M$ by a factor of five only decreases $\kappa(MA)$ by a factor of two.
It is therefore difficult to imagine a system size or sparsity pattern where such an incremental increase in $d$ could produce sufficiently large reductions in $\kappa(MA)$ as to make the procedure worthwhile.

\subsection{Fast Inverse}
The fast inverse method of Tong \emph{et al.}~\cite{tong2021fast} solves linear systems $Ax=b$ where $A$ can be decomposed as
\begin{equation}\label{eq:fastinv-decomp}
	A = \Ab+\As,
\end{equation}
where $\|\Ab\| \gg \|\As\|$.
They then give a QLS algorithm which uses $\Ab^{-1}$ as a preconditioner and solves the system $(I + \Ab^{-1}\As)x=\Ab^{-1}b$ with scaling bounded by $\|\As\|, \|\Ab^{-1}\|,$ and $\|A^{-1}\|$.

This technique is dependent on efficient block-encodings of $\Ab^{-1}$ and $\As$.
An $(\alpha, m, \epsilon)$-block-encoding of the matrix $A$ is given by the unitary $U_A$:
\begin{equation}
	U_A = \begin{bmatrix}
		  	A/\alpha & *\\
		  	* & *
		   \end{bmatrix}
\end{equation}
where $*$ denotes arbitrary matrix blocks, $\alpha$ is a rescaling constant such that $\|U_A\|=1$, and the error $\epsilon$ is bounded by $\|A-\alpha(\langle0^m|\otimes I_n)U_A(|O^m\rangle\otimes I_n)\| \le \epsilon$.
Since the magnitude of $\alpha$ plays a critical role in the scaling of this algorithm, we note that $\|U_A\|=1$ implies that $\alpha \ge \|A\|$.

In order to use $\Ab^{-1}$ as the preconditioner, $\Ab$ must be \emph{fast-invertible}. 
A matrix $M$ is fast-invertible matrix if one can efficiently prepare a $(\Theta(1), m, \epsilon)$-block-encoding $U'_M$ of $M^{-1}$.
This requires access to an oracle for $M^{-1}$, and the number of queries to this oracle in preparing $U'_M$ must be independent of $\kappa(M)$.
For example, if $M$ is normal, and the eigenvalue decomposition $M=VDV^{\dagger}$ gives a $V$ that can be efficiently implemented in a quantum circuit and the elements of $D$ can be accessed through an oracle, then $M$ is fast-invertible.

The fast-inverse QLS algorithm takes as inputs an $(\alpha_s, m_s, 0)$-block-encoding $U_s$ of $\As$, and an $(\alpha'_b, m'_b, 0)$-block-encoding $U'_b$ of $\Ab^{-1}$ implemented via fast-inversion.
They then use a modified version of the quantum singular value transformation~\cite{gily2019quantum} to construct a block encoding of $(\Ab+\As)^{-1}$ with error $\epsilon$ in  
\begin{equation}\label{eq:fastinv-scaling}
	O\left(\frac{\alpha'_b\alpha_s}{\tilde{\sigma}_{\min}}\log \left(\frac{\alpha'_b}{\tilde{\sigma}_{\min}\epsilon}\right)\right)
\end{equation}
applications of $U_s, U'_b$ along with their inverses, controlled versions, and other primitive gates.
Here $\tilde{\sigma}_{\min}$ is a lower bound for the smallest singular value of $I+\Ab^{-1}\As$, i.e. the preconditioned system. 

This approach has the benefit of providing an upper bound on the condition number of the preconditioned matrix, with the downside of needing a decomposition of $A$ that matches a lengthy list of requirements. 
For fracture problems, we have the natural decomposition of
\begin{equation}
 	A = \Delta + A_F,
\end{equation}
where the Laplacian $\Delta$ describes the flow in the absence of fractures, and $A_F$ denotes the fracture matrix.
Fortunately the discretized Laplacian is normal and has a known eigenvalue decomposition~\cite{demmel1997applied}, therefore it meets the  fast-invertible conditions and we may use it as the preconditioner.
However, we have no guarantee that $\|\Delta\| \gg \|A-\Delta\|$, which is required to get good scaling.
Still, we can numerically test the scaling of the algorithm to see how it performs in the absence of performance guarantees. 

The parameters contributing to the performance of this algorithm, Eq.~\ref{eq:fastinv-scaling}, are the block-encoding parameters $\alpha'_b$ and $\alpha_s$, along with a lower bound on the smallest singular value of the preconditioned system $MA$, $\tilde{\sigma}_{\min}$.
In order to assess the potential usefulness of this algorithm for our application, we will explore the most optimistic values for these parameters.
Due to minor technical details, we rescale the entire system by $\|\Delta^{-1}\|$, which gives $\Ab = \Delta\|\Delta^{-1}\|$ and $\alpha'_b$ (the block-encoding parameter for $\Ab^{-1}$) $ = \Theta(1)$.
We also have $\As = (A-\Delta)\|\Delta^{-1}\|$, so $\alpha_s \ge \|A-\Delta\|\|\Delta^{-1}\|$, and $1/\tilde{\sigma}_{\min} \ge \|A^{-1}\Delta\|$.
Therefore the overall scaling for the fast-inverse QLS algorithm is bounded below by
\begin{equation}
	O\left(\frac{\alpha'_b\alpha_s}{\tilde{\sigma}_{\min}}\log \left(\frac{\alpha'_b}{\tilde{\sigma}_{\min}\epsilon}\right)\right) \ge O\left(\|\Delta^{-1}\|\|A-\Delta\|\|A^{-1}\Delta\|\log \left(\|A^{-1}\Delta\|/\epsilon\right)\right).
\end{equation}

\bibliography{references}

\section*{Acknowledgements}

This work was supported by the U.S. Department of Energy through the Los Alamos National Laboratory, LA-UR-22-24431. Los Alamos National Laboratory is operated by Triad National Security, LLC, for the National Nuclear Security Administration of U.S. Department of Energy (Contract No. 89233218CNA000001). DO and JG acknowledge support from Los Alamos National Laboratory's Laboratory Directed Research and Development program through project 20220077ER.

\section*{Author Contributions}

DO and JG conceived the study,  JG conducted the experiments and analysed the results. All authors reviewed the manuscript. 

\section*{Data Availability}

The datasets used and/or analyzed during the current study are available from the corresponding author on reasonable request.

\end{document}